\documentstyle[12pt,aasms4]{article}
\tighten

\newcommand\be{\begin{equation}}
\newcommand\ee{\end{equation}}
\newcommand\jcd{Christensen-Dalsgaard}

\input epsf
\lefthead{Basu and Antia}
\righthead{Effect of peak profile asymmetry on frequencies}
\begin{document}

\title{Effect of asymmetry in peak profiles on solar oscillation frequencies}

\author{Sarbani Basu}
\affil{Institute for Advanced Study, Olden Lane, Princeton NJ 08540,
U. S. A.}
\and
\author{H. M. Antia} 
\affil{Tata Institute of Fundamental Research, 
Homi Bhabha Road, Mumbai 400005, India}

\begin{abstract}
Most helioseismic analyses are based on solar oscillations
frequencies obtained by fitting
symmetric peak profiles to the power spectra. However, it has now been
demonstrated that the peaks are not symmetric. In this work we study
the effects of asymmetry of the peak profiles on the solar oscillations
frequencies of p-modes for low and intermediate degrees. We also
investigate how the resulting shift in frequencies  affects
helioseismic inferences.
\end{abstract}


\section{Introduction}

Accurately measured frequencies of solar oscillations have been
extensively used to infer the properties of solar interior. Most of
the frequency tables available so far (e.g., Hill et al.~1996; Rhodes
et al.~1997) have been obtained by fitting symmetric Lorentzian peak
profiles to the observed power spectra. However, it has been
demonstrated that in general, the peaks in solar oscillation power
spectra are not symmetric (Duvall et al.~1993; Toutain 1993; Nigam \&
Kosovichev 1998; Toutain et al.~1998; Antia \& Basu 1999) and the use
of symmetric profiles may cause the fitted frequency to be shifted
away from the true value. These frequency shifts may affect the
helioseismic inferences obtained from existing frequency tables based
on fits to symmetric peak profiles.

While the asymmetric nature of the peaks in the power spectra is 
well established, what is not known for certain is how much
the frequency shifts caused by fitting the symmetric profiles to the 
peaks affect inferences about the solar interior. 
Toutain et al.~(1998) have studied the effect of asymmetry on the
frequencies of low degree modes and concluded that the inferred sound
speed in the solar core can be significantly affected by the resulting
frequency shifts. However,  they did not include the effects of
asymmetry on intermediate degree modes, which are also needed
for inferring conditions in solar interior accurately. 
To get a proper idea of this
effect it is necessary to include the effect of asymmetry on the
intermediate degree modes also. 
Christensen-Dalsgaard et al.~(1998)
on the other hand concluded that the 
frequency shifts due to asymmetry in peak profiles should not cause any
significant change in inversion results. 
This conclusion was reinforced by
the inverse analyses carried out by Rabello-Soares et al.~(1999).
Their results, however, are based on artificial data, where they have
assumed that the dimensionless asymmetry parameter characterizing the
asymmetry of peak profiles is a function of frequency alone.
Thus these results need to be checked against those obtained from
real spectra for solar oscillations.

Apart from low degree modes, the frequency shifts due to asymmetry
in peak profiles are also found to be significant in high degree modes
obtained from ring diagram analysis (Antia \& Basu 1999). Thus
it would be interesting to study how the use of
asymmetric profiles affects the frequencies of intermediate degree
modes obtained from full-disk observations. In this work we use data from the
Global Oscillations Network Group (GONG) to study the effects of
peak-profile asymmetry on frequencies of p-modes with degree $0\le\ell\le200$
using the rotationally corrected, $m$-averaged power spectra, $m$ being the 
azimuthal order of the mode.

The rest of the paper is organized as follows:  the basic technique
used to determine solar oscillations frequencies using asymmetric
peak profiles is
described in \S~2. The resulting frequency shifts are described in \S~3,
while the effects on
helioseismic inferences are described in \S~4.
The conclusions from our study are summarized in \S~5. 

\section{The technique}

We use  GONG power spectra to determine the frequencies of solar
oscillations. The GONG project determines the frequencies of modes
with different values of $n$ (radial order), $\ell$ (degree) and $m$ (azimuthal order)
by fitting
symmetric Lorentzian peak profiles to spectra for individual values
of $\ell,m$ (Hill et al.~1996). 
The mean frequency of a  multiplet for a given ($n,\ell$) pair is then calculated
by fitting the frequencies of all modes with same $n$ and
$\ell$   but different values of $m$ to
polynomials in $m$.
Since the asymmetry in peak profiles is relatively small, it is
difficult to distinguish between fits to symmetric and asymmetric
profiles in spectra for individual $\ell,m$.
To improve 
statistics we use the $m$-averaged spectra obtained by taking a sum
over the azimuthal order $m$ for each $\ell$. Since in this work we
are only interested in the mean frequency for each $n,\ell$ multiplet,
we correct for the rotational splitting by shifting the spectrum for
each $m$ by the approximately known rotational splitting before
summing. Such spectra are available from the GONG project (Pohl \&
Anderson 1998) for each of the GONG months 1--35. Each GONG `month'
covers a period of 36 days. 
In order to improve statistics still further, we have summed the
spectra for different months.
We can, in principle,
sum all 35  spectra, but in view of
the solar cycle variation in frequencies this may not be advisable. As
a result we have taken sum over 16 months from month 7 to 22 (9
December 1995 to 6 July 1997), which is the period when the solar
activity was close to minimum and there is little change in
frequencies during this period. Most of our results about asymmetry in
peak profiles have been obtained from this spectrum. 

To determine the frequencies and other mode parameters from the
power spectra, we fit a model of the form
\be
P(\ell,\nu)=
\sum_i\left(\exp(A_i)(S^2+(1+Sx_i)^2)\over
x_i^2+1\right)+B_1+B_2(\nu-\nu_c),
\label{eq:model}
\ee
where $x_i=(\nu-\nu_i)/ w_i$ and the summation is carried over all
peaks in the fitting interval. If there are $N$ peaks in the fitting
interval then the $3N+3$ parameters $A_i, \nu_i,w_i, S, B_1$ and $B_2$
are determined by fitting a section of the spectra at constant $\ell$
using a maximum likelihood approach (Anderson, Duvall \&
Jefferies~1990). In Eq.~(\ref{eq:model}),
 $\nu_c$ is the central value of $\nu$ in the
fitting interval and $\exp(A_i)$ is the peak power in the mode,
$\nu_i$ is the mean frequency of the corresponding peak, $w_i$ is the
half-width. The terms involving $B_1,B_2$ define the background power,
which is assumed to be linear in $\nu$. $S$ is a parameter that
controls the asymmetry, and the form of asymmetry is the same as that
prescribed by Nigam \& Kosovichev~(1998). This parameter is positive
for positive asymmetry, i.e., more power on the higher frequency side
of the peak, and negative for negative asymmetry. By setting $S=0$ we
can fit symmetric Lorentzian profiles. We have assumed that $S$ has
the same value for all peaks in the fitting interval. This may not be
strictly true but the variation in $S$ is not very large between
neighboring peaks and for simplicity we neglect its variation between
peaks in the fitting interval. This improves the convergence of
fitting procedure.
Even then, inclusion of asymmetry parameter in the fits reduces the
number of modes that are successfully fitted. This could be due to
some cross-correlation between $S$ and other parameters of the model,
particularly, the background.

We fit each mode separately by using the portion of power spectrum
extending halfway to the adjoining modes. Apart from the target mode
there  are other peaks in the spectra arising due to leaks 
from neighboring $\ell$ and $n$ values. We include all leaks 
from modes for which $\ell$ differs by at most 3 
from those of the target mode, provided they occur within the
fitting interval.
Although all these
peaks are fitted to obtain a good fit to the observed spectra, we
ultimately use only the parameters obtained for the target peak and
ignore the leaks. 

The use of rotationally corrected, $m$-averaged spectra may introduce
some systematic errors in the frequency due to incorrect even order
splitting coefficients used in constructing the spectra. These
coefficients are assumed to be zero while constructing the GONG
$m$-averaged spectra. The non-zero values of these coefficients will
introduce a small shift in the frequencies, but that is not relevant
in the current work as we are only interested in the effect of
asymmetry on the frequencies. The splitting coefficients will affect
the fits to both symmetric and asymmetric spectra equally and its
effect will cancel out when we take frequency differences between the
two fits. Nevertheless, we can estimate this systematic error by taking the
difference between frequencies fitted by us and those obtained by
the GONG project using individual $\ell,m$ spectra.

\section{The frequency shifts}

We follow the procedure outlined in \S~2 to fit the model given by
Eq.~(\ref{eq:model}) to suitable regions of the spectra obtained by summing over
the 16 spectra for GONG months 7--22 in order to
determine the mode parameters.
Although other parameters may be of interest, in this work we only
concentrate on the frequencies $\nu_i$ and the asymmetry parameter
$S$. We fit both symmetric and asymmetric peak profiles to the spectra
for studying the shift in frequency arising due to asymmetry in peak
profiles. Fig.~1 shows a fit to the $\ell=100$ spectrum using both
symmetric and asymmetric profiles. It is clear from the figure that
asymmetric profile gives a better fit to the observed spectra and it
is probably desirable to use asymmetric profiles to determine
frequencies. For comparison this figure also shows a fit to the $\ell=0$
power spectrum over a  similar frequency range. The $\ell=0$ spectrum
does not involve any sum over spectra for different $m$.
It is clear that the spectrum in this case is too noisy to
distinguish between the two fits. This shows why we have used the
$m$-averaged spectra in this study ---  
summing over spectra for all values of $m$ increases the signal to
noise ratio.

Fig.~2 shows the asymmetry parameter $S$ for the modes.
It is clear that this parameter is significant at  frequencies
around 2--2.5 mHz. This parameter is negative for all modes and hence
there is more power on the low frequency side of the peak. The
magnitude of $S$ is similar to what has been found by Toutain et
al.~(1998) for low degree modes and by Antia \& Basu~(1999) for high
degree modes. The variation of $S$ with frequency is somewhat
different from what was found at high degree using the ring diagram
analysis. This probably implies that apart from frequency the asymmetry
may also depend on $\ell$.

The shift in frequencies that result from using asymmetric
peak profiles rather than the standard Lorentzian profiles are
shown in Fig.~3.
These frequency shifts are positive, i.e.,  frequencies tend to
increase when asymmetric profiles are used. These frequency shifts are
clearly larger than the estimated errors
and hence this effect must be included in helioseismic analysis.
However, the frequency shift appears to be a function
predominantly of frequency and is only weakly dependent on $\ell$.
If this is true then the difference may be accounted for by
the surface term in helioseismic inversions (\jcd\ et al.~1998).
In order to check for any depth dependence we
also show in Fig.~3 the frequency difference as a function 
of the lower turning point ($r_t$) for the mode. It is clear
that there is a weak dependence of frequency difference with $r_t$,
and there appears to be a change around the base of the convection
zone, which is located at a radial distance of $0.713R_\odot$
(Christensen-Dalsgaard, Gough \& Thompson~1991; Basu 1998).
Thus we may expect some change in the inferred properties of solar
interior when asymmetry in peak profiles is incorporated.
In the next section we investigate the effect of these frequency
shifts on various helioseismic inferences. Similar frequency
shifts have been obtained for spectra from different time periods.

As mentioned in Section~2, there may be some systematic errors
introduced by using the $m$-averaged spectra for determining the
frequencies. In order to estimate this error we repeat the
calculations for the summed spectra from the GONG months 4--14
(23 August, 1995 to 21 September, 1996) using
symmetric peak profiles and compare the results with the mean
frequencies determined from fitting the individual $n,\ell,m$ modes
by the GONG project.
The frequency difference is shown in Fig.~4. It is clear that the
systematic errors are of order of $0.01\mu$Hz, which is much smaller
than the frequency shift due to asymmetry in peak profiles. Moreover,
as mentioned earlier these systematic errors will cancel when we take
the difference between frequencies from symmetric and asymmetric peak
profiles. 

\section{Effect of asymmetry on structure inversion results}

To investigate the effect of frequency shift due to asymmetric
peak profiles on helioseismic inversions
for solar structure, we first try the asymptotic inversion technique
(\jcd, Gough \& Thompson 1989). For this purpose the frequency difference
is expressed as
\be
S(w){\delta\omega\over\omega}=H_1(w)+H_2(\omega)
\ee
where $w=\omega/(\ell+1/2)$ and
\be
S(w)=\int_{r_t}^{R_\odot} \left(1-{c^2\over w^2r^2}\right)^{-1/2}
{dr\over c}.
\ee
Here, the function $H_1(w)$ contains information about the variation
of sound speed with depth and can be inverted to obtain the sound speed,
while $H_2(\omega)$ represents the effect of differences in surface
layers. This analysis can be applied to the frequency shifts shown in
Fig.~3 to obtain the error introduced in inversion results due to
asymmetry in peak profiles. The results are shown in Fig.~5.
It may be noted that both $H_1(w)$ and
$H_2(\omega)$ are comparable in magnitude and hence the frequency shift
can not entirely be considered as surface effects and we would expect
some change in inferred solar structure also. Fig.~6 shows the
inferred relative difference in sound speed due to the frequency shifts
shown in Fig.~3 and it is clear that the difference 
is fairly small, being comparable to
the estimated errors in inversions. There is a very small hump
near the base of the convection zone, which may give some difference
in the estimated depth of the convection zone or the extent of overshoot
below the convection zone. The small difference in the
convection zone may account for some of the observed difference between
the Sun (as inferred using fits to symmetric profiles) and standard
solar models.

Instead of asymptotic inversion we can perform non-asymptotic
inversions using the Regularized Least Squares (RLS) method
(Antia 1996) or Subtractive Optimally Localized Averages (SOLA) technique
(Basu et al.~1996). These results are also shown in Fig.~6 and are
similar to those obtained using asymptotic inversion technique.
It is clear from all these results that the frequency shifts due to
asymmetry in peak profiles do not affect the structure inversion results
significantly. 
The difference $\delta c/c$ in the core is much less than
what was found by Toutain et al.~(1998) who found relative sound speed
differences exceeding $0.002$. Since the small change appears to
manifest as hump around the base of the convection zone,
in the next two subsections we
investigate the effect of this frequency shifts on the inferred depth
of the convection zone and the extent of overshoot below the convection
zone. There is also a small dip near $r=0.5R_\odot$, which can be seen
in frequency difference shown in Fig.~3 too.
This dip is comparable to error estimates in the individual modes,
though after averaging over neighboring modes it may appear to be
somewhat significant. The origin of this dip is not clear and it may be a
numerical artifact arising from some correlations in spectra or
between different parameters of model fitted. This dip is present in
results obtained from most of the spectra that we have fitted.
However, in averaged spectra from GONG months 24 to 35 and months 32 to 35
this dip can barely be seen.

\subsection{Depth of the convection zone}

Using solar p-mode frequencies it is possible to determine the
depth of the convection zone quite precisely (\jcd, Gough \& Thompson
1991; Basu \& Antia 1997) and it would be interesting to check if this
depth is affected by the frequency shifts resulting from asymmetry in
peak profiles. We follow the approach used by Basu \& Antia (1997) to
determine the depth of the convection zone using the frequencies
as obtained by fitting both
symmetric and asymmetric profiles. We use the same set of
reference models to determine the depth of convection zone using the
two sets of frequencies. The fits to symmetric profiles yield the
position of the base of the convection zone at $(0.71336\pm
0.00004)R_\odot$, while the use of frequencies obtained by fitting asymmetric
profiles yield a value $(0.71344\pm 0.00005)R_\odot$. Thus there is a
marginal decrease in the inferred depth of the convection zone by
$0.00008R_\odot=56$ km due to asymmetry in peak profiles, which is
comparable to the error estimates. However, these error estimates do
not include systematic errors as discussed by Basu \& Antia~(1997) and
Basu (1998), which are an order of magnitude larger. Thus the
difference arising due to asymmetry is essentially insignificant. 
Note that the error estimate is slightly larger for frequencies
obtained from asymmetric profiles since in that case the number of
modes successfully fitted is somewhat smaller.

\subsection{Overshoot below the convection zone}

Apart from the depth of the convection zone, it is also possible to
estimate the extent of overshoot below the solar convection zone from
the measured frequencies of solar oscillations (Gough 1990;
Monteiro, \jcd\ \& Thompson~1994;
Basu, Antia \& Narasimha 1994; Basu 1997). This measurement is
obtained from a characteristic oscillatory component in frequencies of
oscillations as a function of $n$, which is introduced by steep
changes in derivatives of the sound speed near the base of the
convection zone, where the temperature gradient changes from adiabatic
value inside the convection zone to the radiative gradient in the
radiative interior. The amplitude of the oscillatory component is a
measure of the extent of overshoot, while the `frequency' of
oscillatory component gives the acoustic depth $\tau$ of the
discontinuity in derivatives of sound speed.
This oscillatory signal can be magnified by taking the fourth difference of
the frequencies as a function of $n$. We follow the approach
used by Basu~(1997) to determine the amplitude and `frequency' of
oscillatory component in frequencies.

The results obtained
using the two sets of frequencies obtained from fits to symmetric and
asymmetric profiles are shown in Fig.~7.  
These can be compared with earlier results obtained using the GONG
data from individual $\ell,n,m$ modes, as well as those from the
MDI data for the first 144 days of its operation (Rhodes et al.~1997).
It is interesting to note that
the results obtained from fits to asymmetric profiles are closer to
those obtained from individual $\ell,n,m$ modes which were fitted to
a symmetric profile. It is possible that this is a coincidence where
systematic errors due to use of $m$-averaged spectra is cancelled by
the frequency shift due to asymmetry. However, the results from MDI
data which also employ symmetric profiles is close to what we find
using symmetric profiles. It appears that the abnormally low
amplitude obtained from MDI data, which is smaller than the amplitude
in a model without overshoot, might be due to use of symmetric profiles
in fitting the power spectra.

\section{Conclusions}

Using the rotationally corrected, $m$-averaged spectra of solar
oscillations obtained by the   GONG network
we have determined
the frequencies of solar oscillations for degree $0\le\ell\le200$. 
The use
of asymmetric peak profiles improves the fit to observed spectra and
the frequencies are increased as compared to those obtained when
symmetric profiles are used. This frequency shift of about $0.2\;\mu$Hz
is larger than the estimated errors in fitted frequency and thus
could affect results of
helioseismic analyses. However, we find that this frequency shift is partly
a function of frequency alone and its effect on helioseismic
inferences is generally smaller than other systematic errors. We
have confirmed this by inverting the frequency differences to estimate
the error in sound speed caused by asymmetry in peak profiles.

We have also investigated how  the frequency shifts affect
results about the depth of the convection zone and the extent of
overshoot below the convection zone.
The inferred depth of the convection zone is
reduced by about 56 km, when the effect of asymmetry is included.
Similarly, the amplitude of oscillatory component in frequencies
increases when asymmetric profiles are used. However, this increase
does not change existing limits on the extent of overshoot
below the solar convection zone (Monteiro et al.~1994; Basu 1997)
since the resulting amplitude is comparable to that obtained
from a solar model without overshoot. In fact, the resulting amplitude
using asymmetric peak profiles is similar to what is found from the
GONG data from individual $\ell,n,m$ modes, which was used in obtaining
earlier limits.

In this work we have investigated the effect of asymmetry in peak
profile on mean frequencies only. In principle, the frequency splittings
may also be affected by asymmetry. Basu \& Antia~(1999) have studied
the effect of asymmetry on the ring diagram analysis of the large
scale flows. They find that the asymmetry in peak profiles does not
affect the inferred velocity field significantly.
The changes in inferred flow velocities due to asymmetry of peaks 
are   equivalent to  changes in odd frequency splitting coefficients
in global p-modes, thus it is possible that the splittings do not change
significantly. 
This may be expected as to a first approximation
asymmetry will shift the frequencies of all modes in a
multiplet for given $n,\ell$ by the same amount and hence the splittings
may not be affected. 
\jcd\ et al.~(1998) have also argued that the effect of asymmetry in peak
profile will not significantly affect the odd splitting coefficients
which are useful in determining the rotation rate in solar interior.
However, the even splitting coefficients which are
determined by
aspherical distortions may be affected by asymmetry in peak profiles.
Clearly, more work is required to investigate the effect of asymmetry
on splitting coefficients. With availability of better data and better
understanding of asymmetry it may be possible to fit asymmetric profiles
to find the splitting coefficients in addition to the mean frequencies
studied in this work.

\acknowledgments

This work utilizes data obtained by the Global Oscillation Network
Group (GONG) project, managed by the National Solar Observatory, a
Division of the National Optical Astronomy Observatories, which is
operated by AURA, Inc. under a cooperative agreement with the National
Science Foundation. The data were acquired by instruments operated by
the Big Bear Solar Observatory, High Altitude Observatory, Learmonth
Solar Observatory, Udaipur Solar Observatory, Instituto de Astrofisico
de Canarias, and Cerro Tololo Interamerican Observatory. 
This work also utilizes data  from the Solar Oscillations
Investigation / Michelson Doppler Imager (SOI/MDI) on the Solar
and Heliospheric Observatory (SOHO).  SOHO is a project of
international cooperation between ESA and NASA.

\begin{figure}
\plottwo{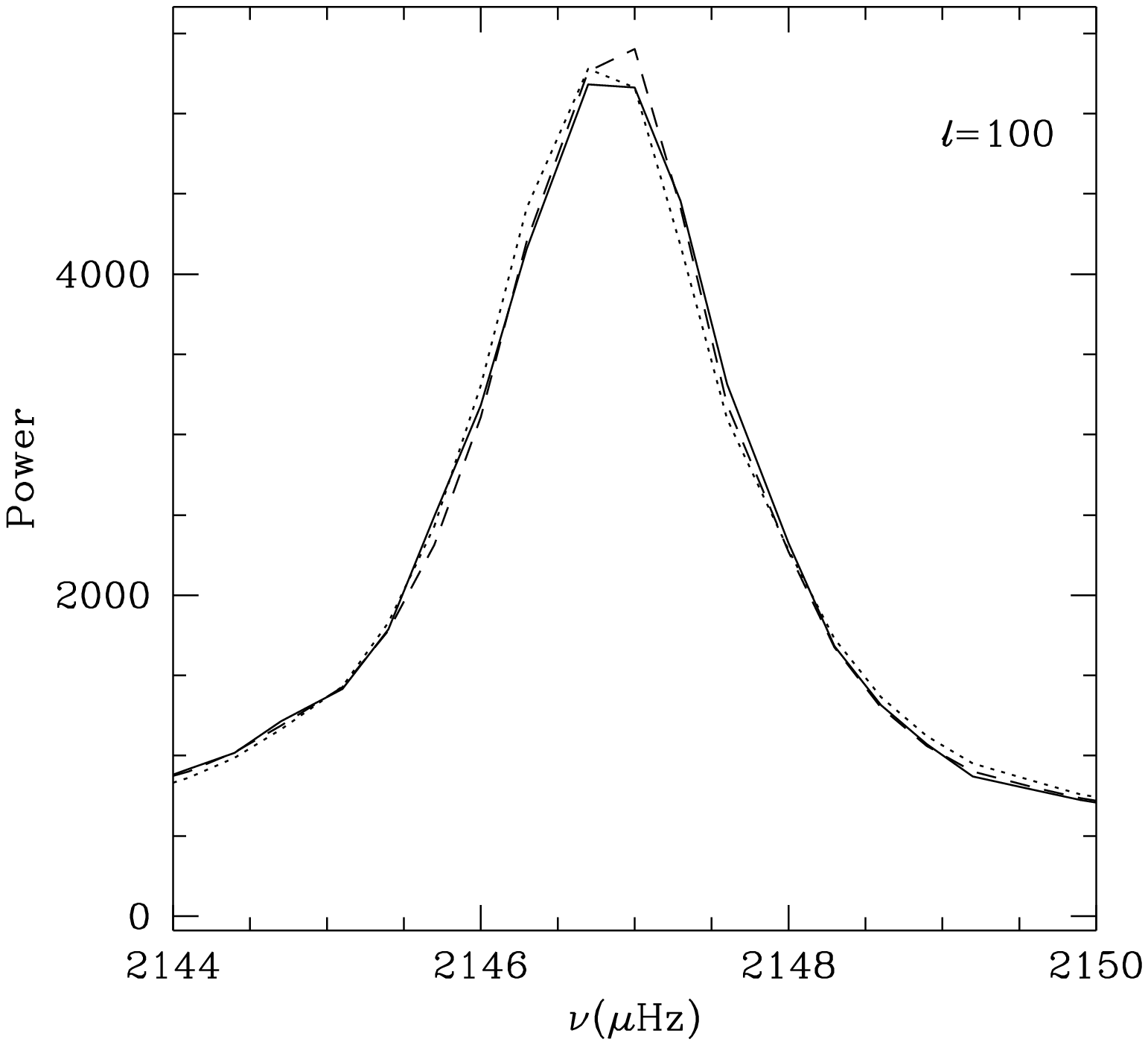}{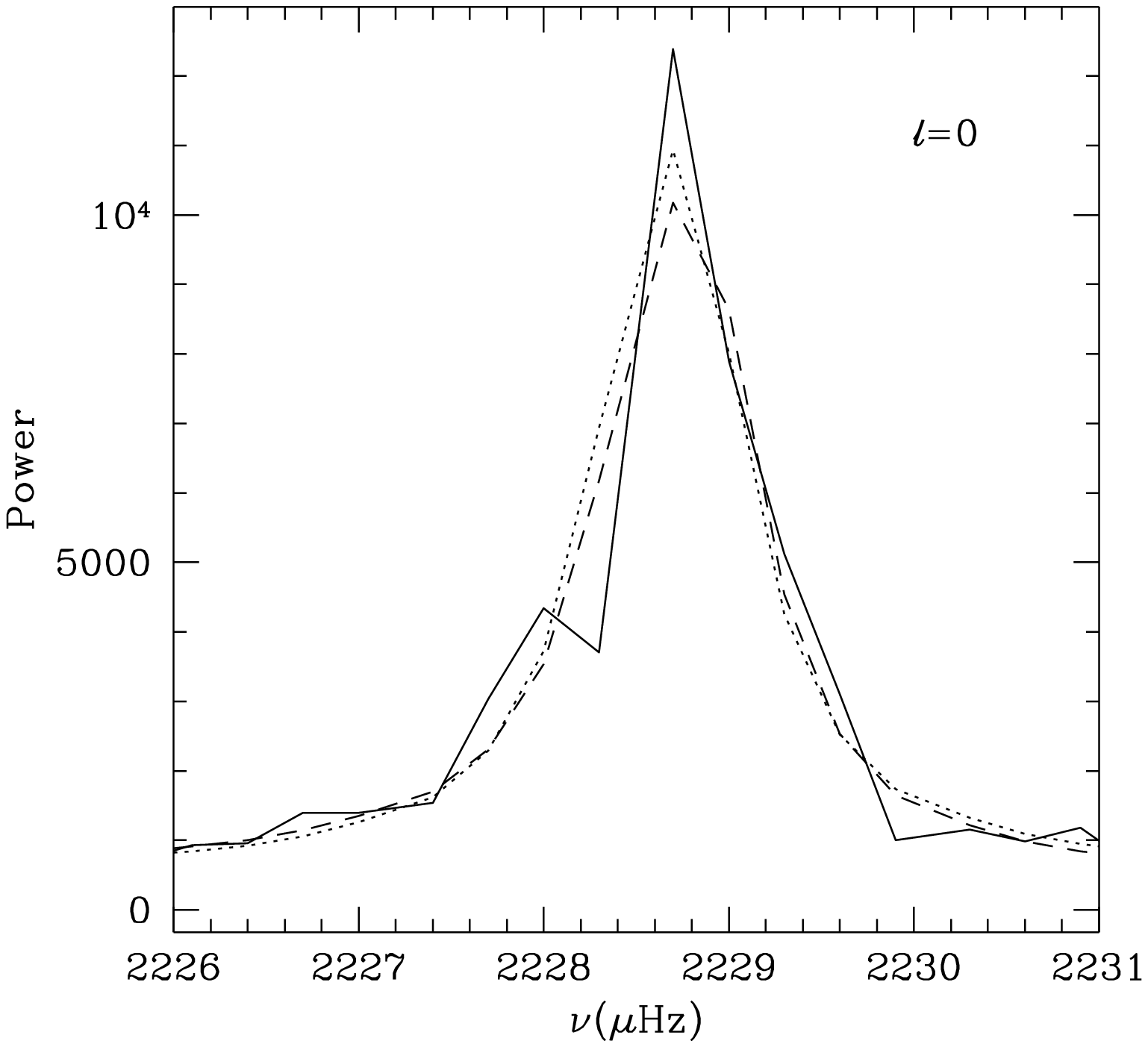}
\figcaption{Fits to power spectra for $\ell=100$, $n=3$ and
$\ell=0$, $n=15$ mode obtained
using symmetric and asymmetric peak profiles. The continuous line
shows the observed power spectra,  the dotted line shows the
fit using symmetric profile and the dashed line shows the fit using
asymmetric profile.
}
\end{figure}

\begin{figure}
\plotone{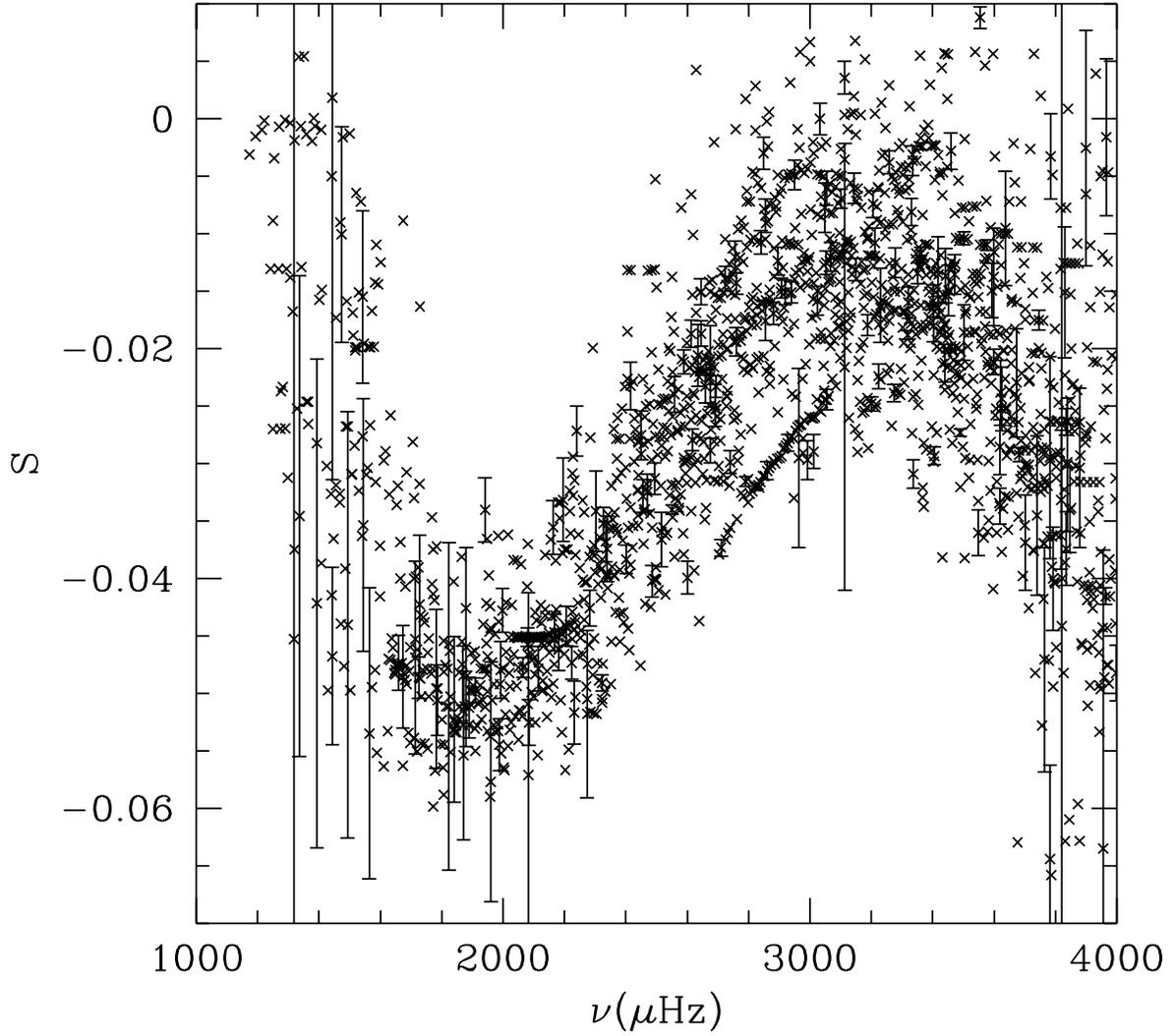}
\figcaption{The asymmetry parameter $S$ for the fits to
summed spectra for GONG months 7--22. For clarity error-bars are shown
only for a few points.
}
\end{figure}

\begin{figure}
\plottwo{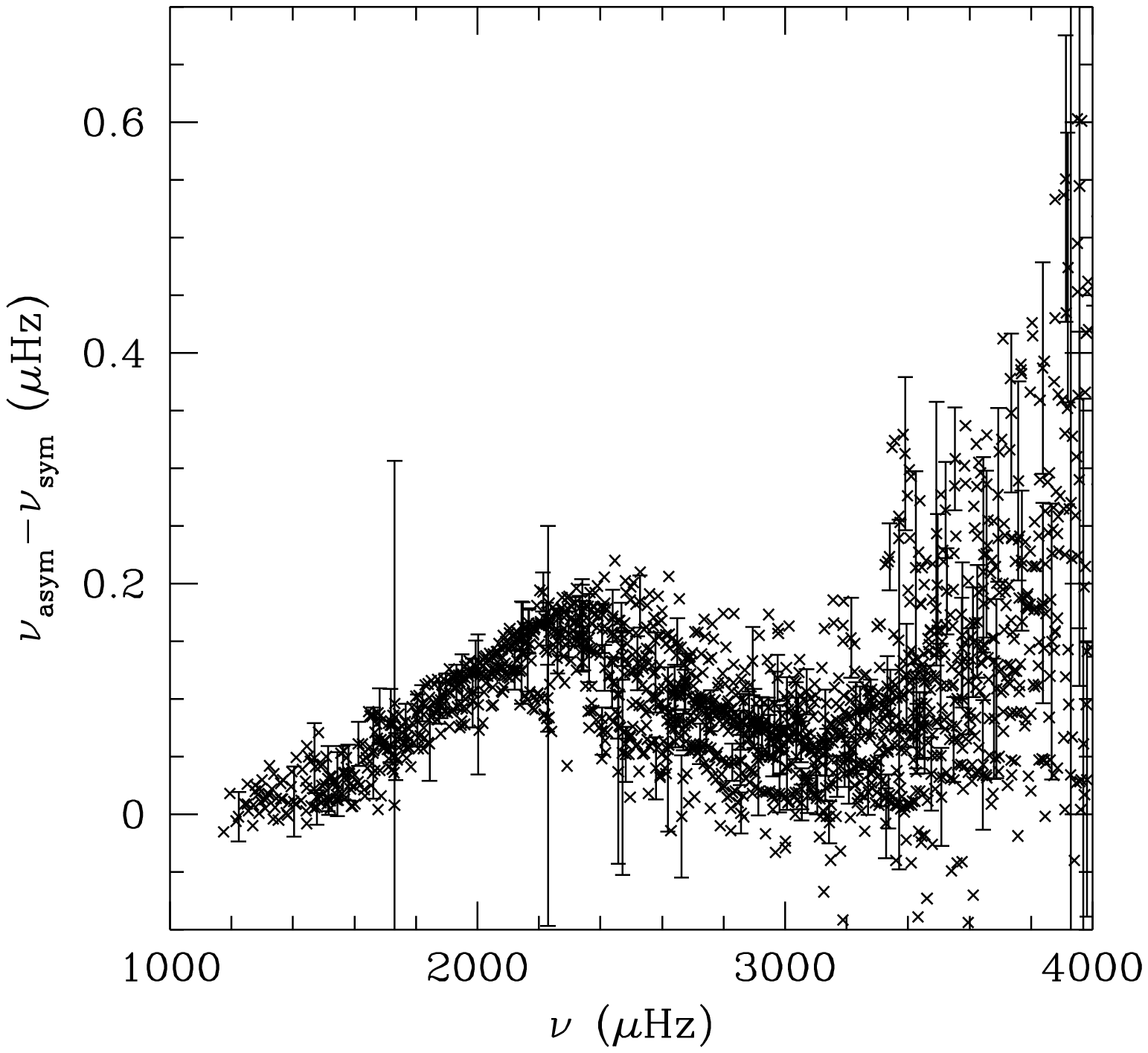}{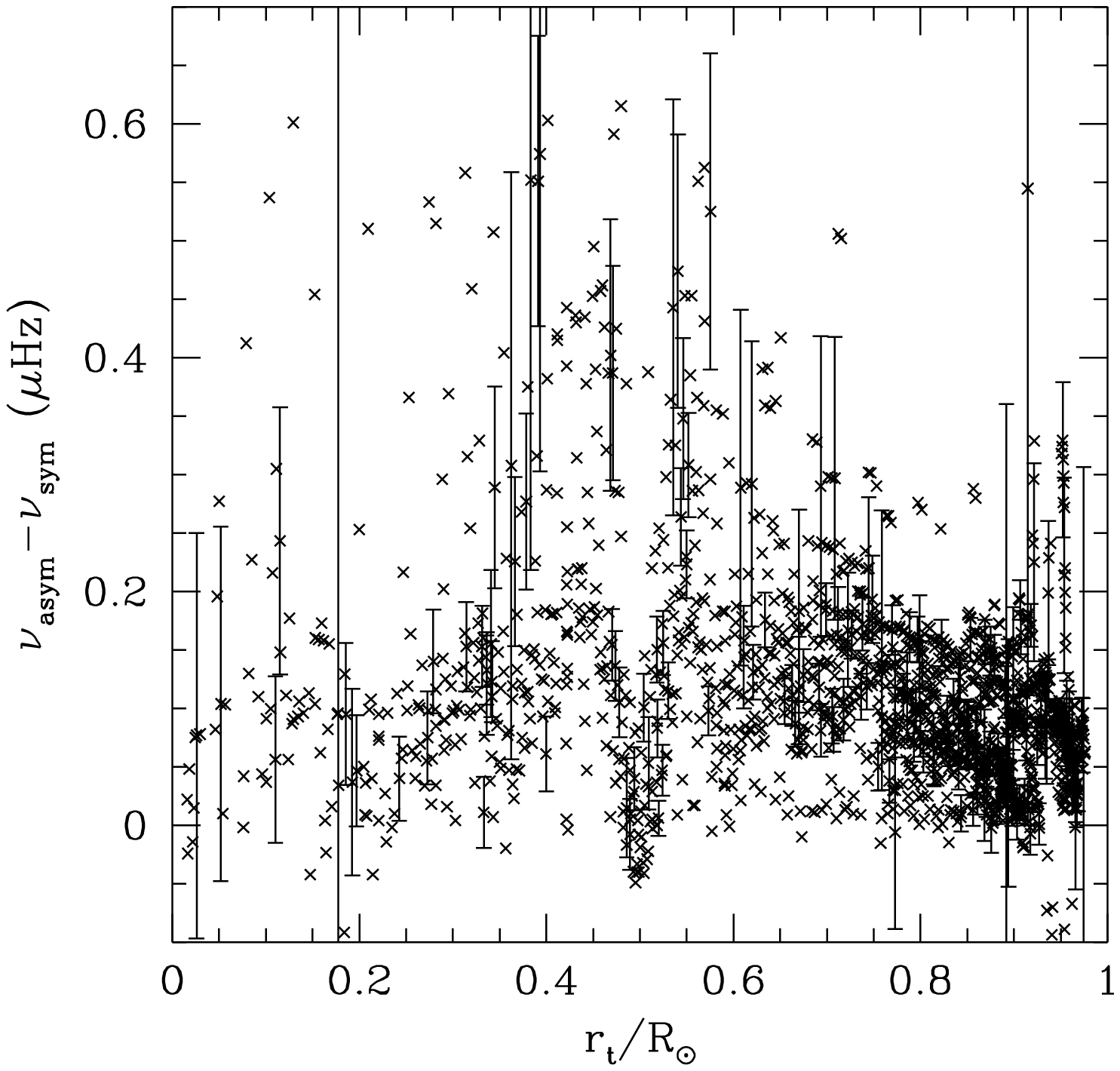}
\figcaption{The frequency shift due to asymmetry in peak profiles
for fits to summed spectra for GONG months 7--22.
For clarity error-bars are shown only for a few points.
}
\end{figure}

\begin{figure}
\plotone{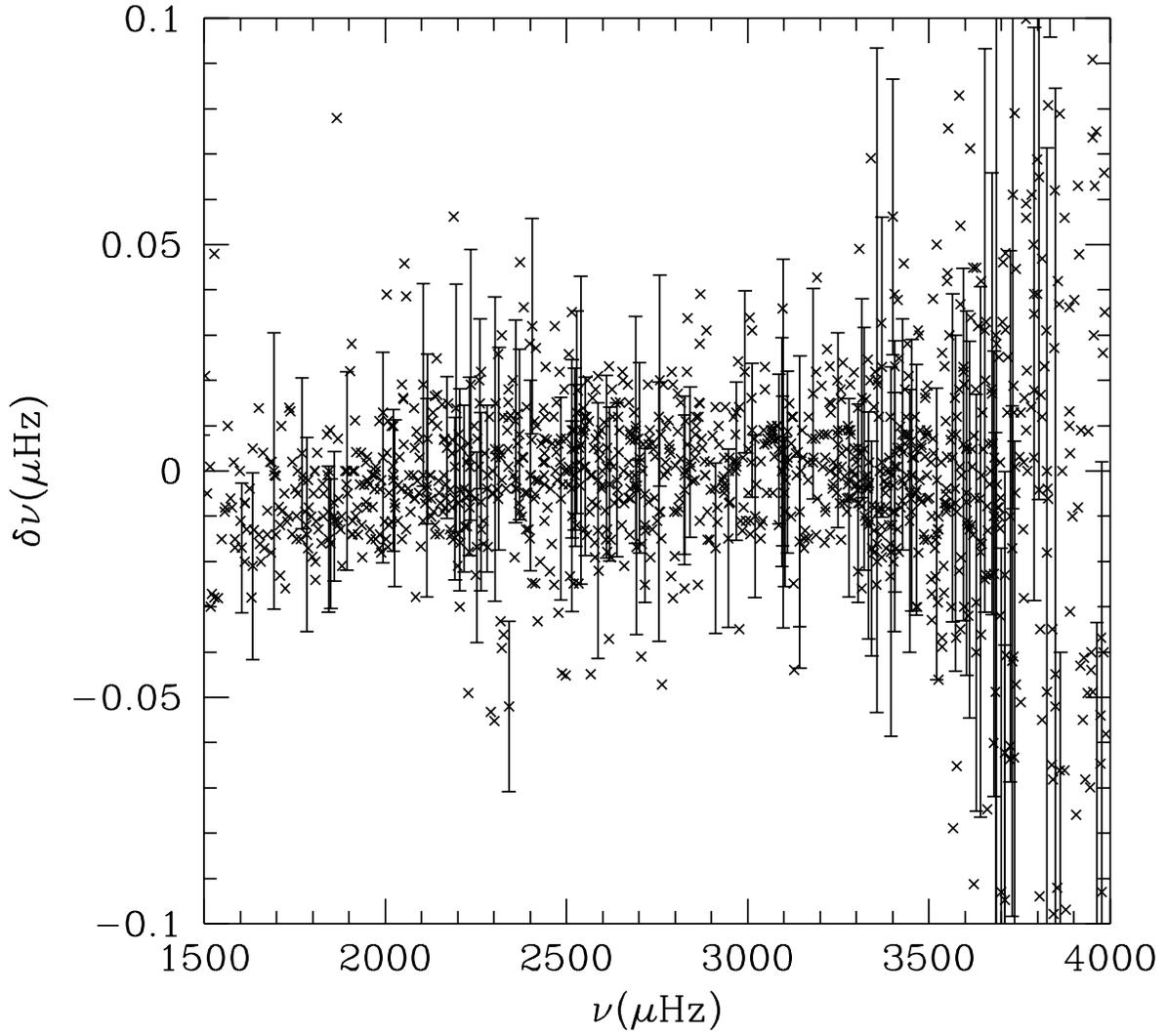}
\figcaption{The frequency difference between the fit to symmetric profiles
for the summed $m$-averaged spectra and those determined by the GONG
project using fits to individual $n,\ell,m$ modes for the GONG months
4--14.
}
\end{figure}

\begin{figure}
\plottwo{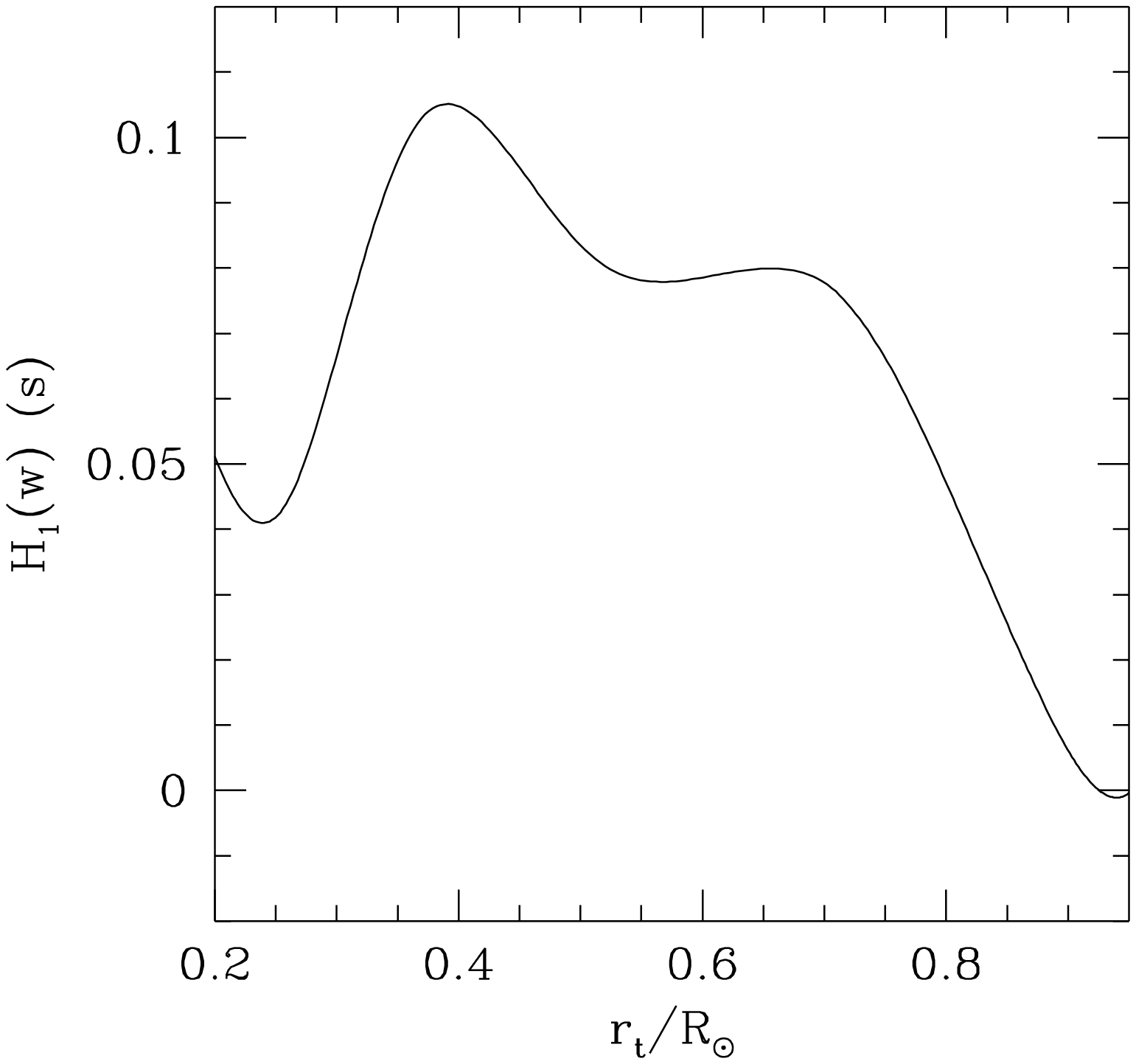}{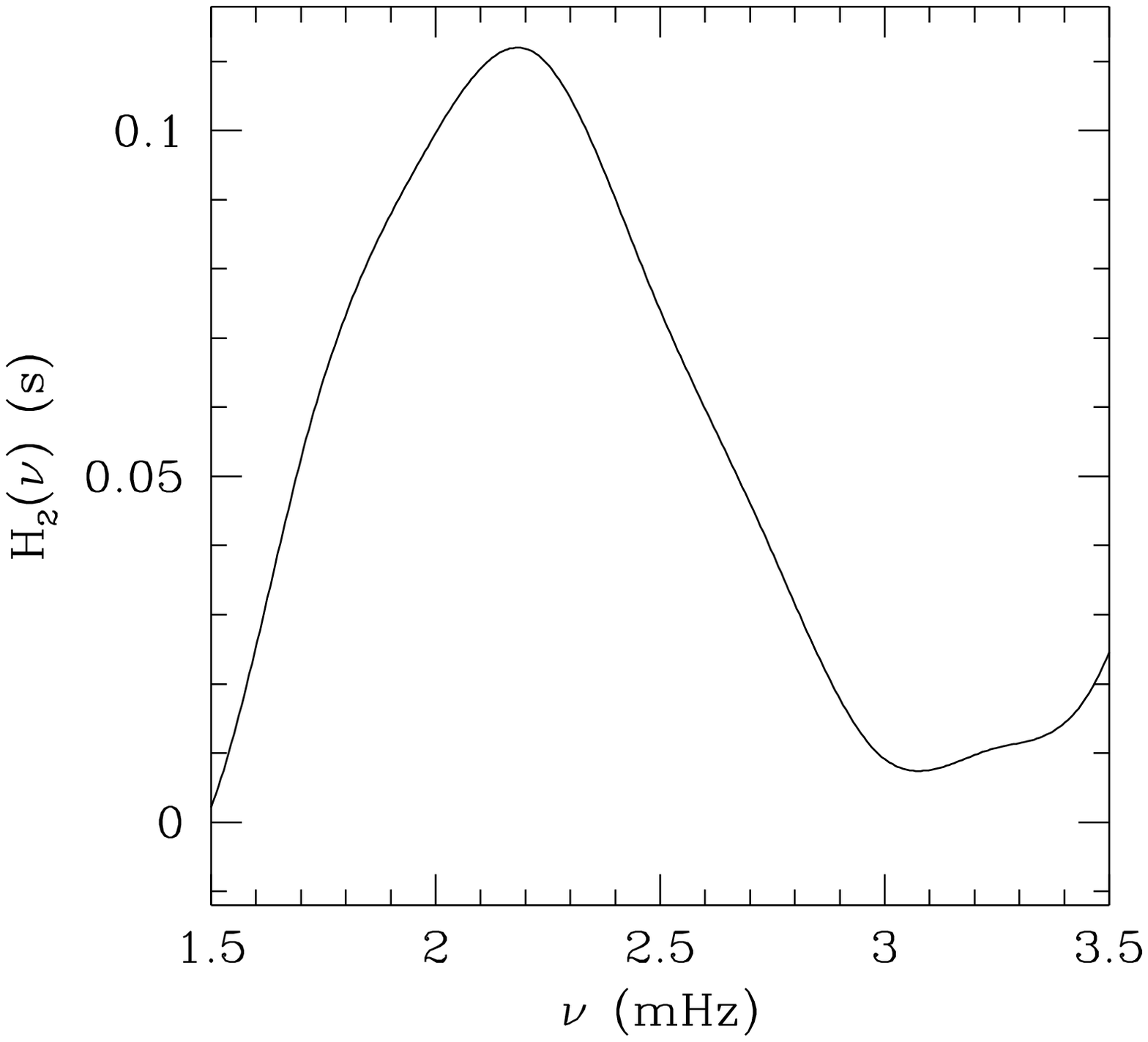}
\figcaption{The functions $H_1(w)$ and $H_2(\nu)$ resulting from the
asymptotic fit to the frequency shifts shown in Fig.~3.
}
\end{figure}

\begin{figure}
\plotone{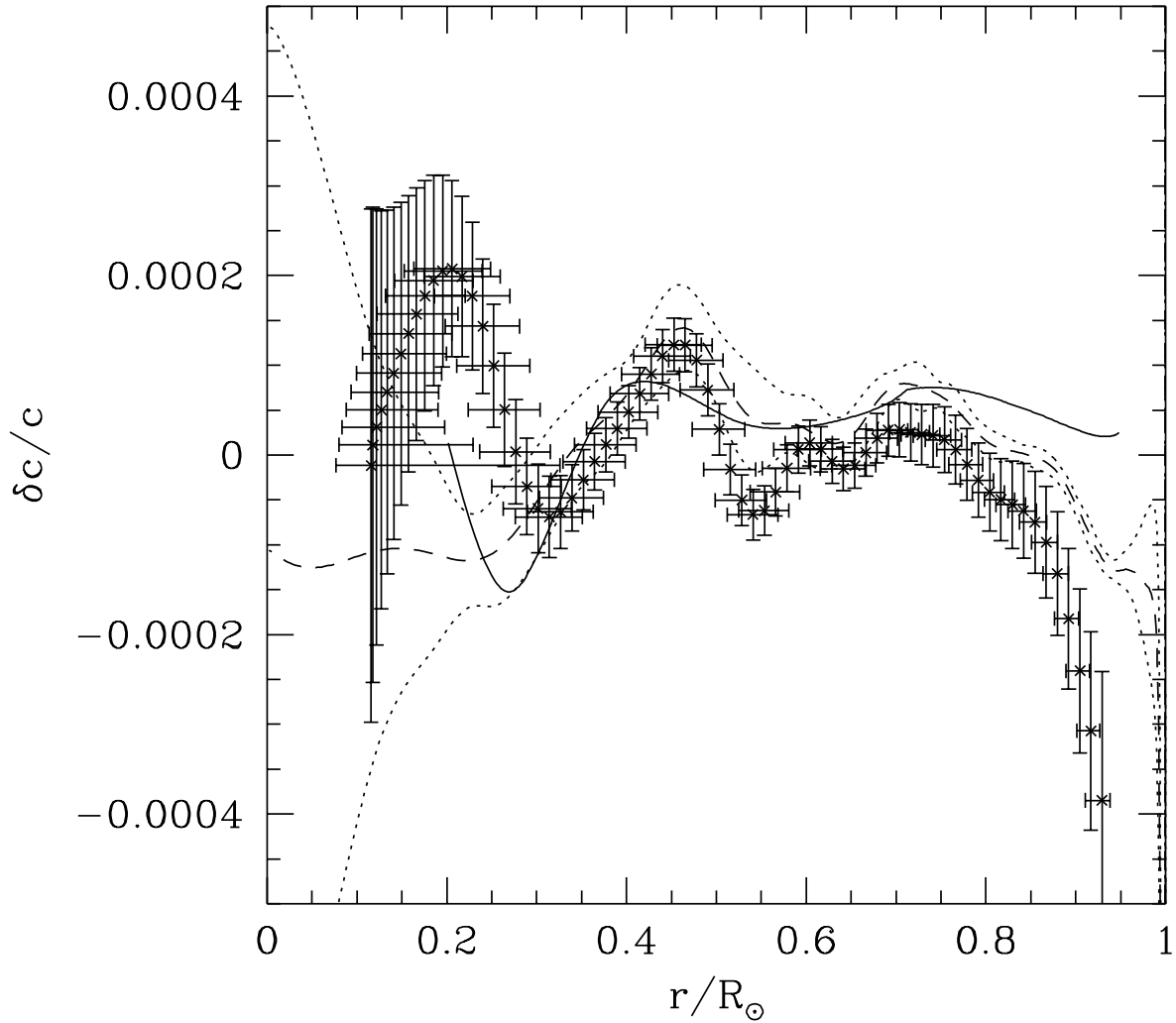}
\figcaption{The relative sound speed difference inferred by various
inversion techniques from the frequency shifts shown in Fig.~3.
These represent the error introduced in helioseismic inversions
due to use of symmetric profiles. The continuous line represents
the results obtained using asymptotic inversion, while dashed line
represents that obtained by RLS technique for nonasymptotic inversion,
with the dotted lines giving
the $1\sigma$ error estimates. The points with error bars represent the
results obtained using OLA technique.
}
\end{figure}

\begin{figure}
\plotone{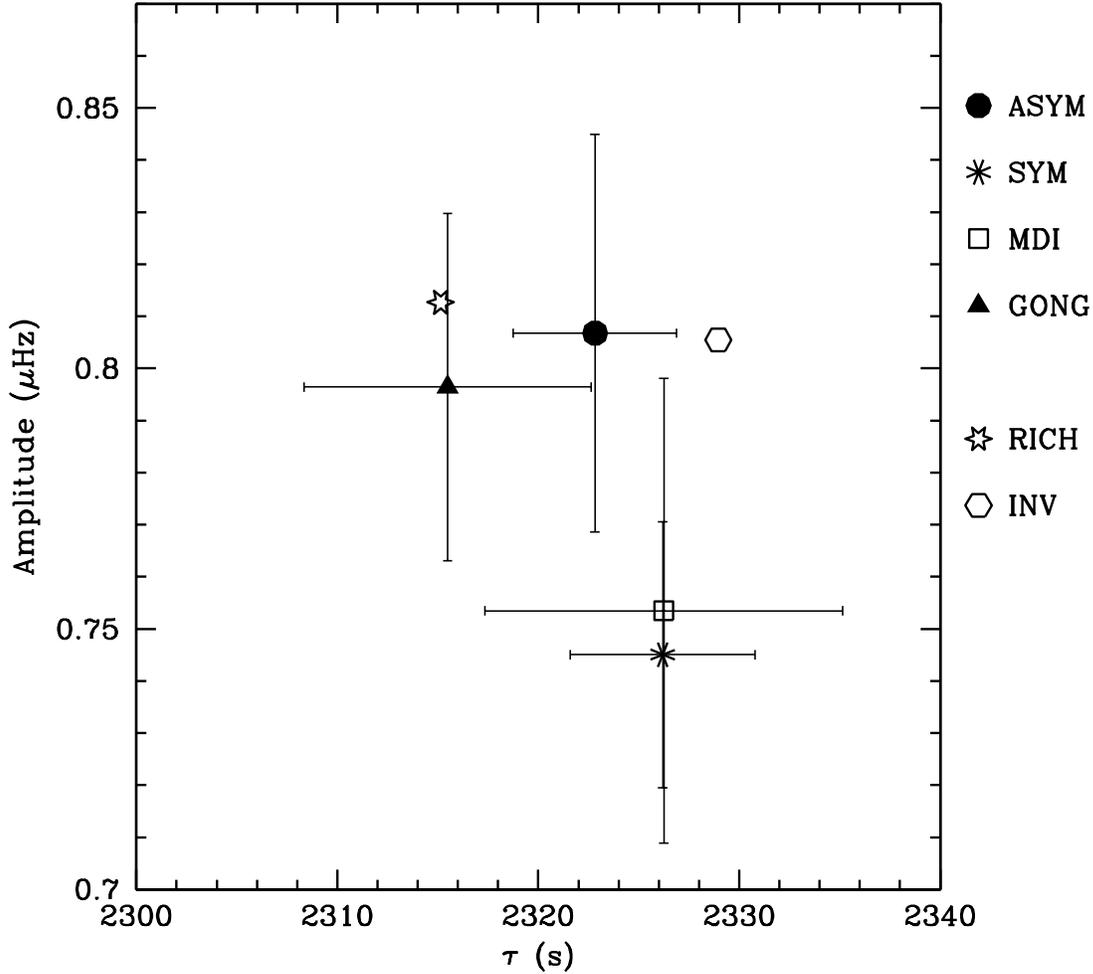}
\figcaption{The amplitude of the oscillatory signal in the
fourth difference of the frequencies plotted as a function of the
`frequency' of the signal. ASYM refers to the frequencies
obtained by fitting asymmetric profiles to the peaks, while
SYM is from frequencies obtained with symmetric
profiles for the GONG months 7--22 summed spectra.
MDI is the results for the MDI 144 day data and GONG
is the result obtained from GONG months 4--10 data where
the frequencies were obtained for each individual $m$-peak.
RICH is for a model using the composition profiles from Model 5 of
Richard et al.~(1996) which includes rotational mixing of elements
below the base of the convection zone,
while INV is that for a model constructed 
with the composition profile obtained from inversions
(Antia \& Chitre 1998). Neither model has overshoot below the
base of the convection zone. 
}
\end{figure}

\end{document}